\documentclass{IEEEtran}
\usepackage{graphicx}
\usepackage{amsmath}
\usepackage{multicol}
\usepackage{multirow}
\usepackage{stfloats}
\usepackage{booktabs}
\usepackage{mathrsfs}
\usepackage{enumerate}
\usepackage{amssymb}
\usepackage{algorithm}
\usepackage{array}
\usepackage{underscore}
\usepackage{url}
\usepackage{hyperref}
\hypersetup{hidelinks}
\usepackage{flushend}

\usepackage{algorithmic}
\usepackage[square, comma, sort&compress, numbers]{natbib}
\usepackage{graphicx}
\usepackage{epstopdf}
\usepackage{caption}
\usepackage{diagbox}

\usepackage{caption}

\usepackage{subfigure}
\usepackage{color}

\usepackage{fancyhdr}

\makeatletter
\newcommand{\Rmnum}[1]{\expandafter\@slowromancap\romannumeral #1@}
\makeatother

\usepackage{array}  \newcommand{\PreserveBackslash}[1]{\let\temp=\\#1\let\\=\temp}  \newcolumntype{C}[1]{>{\PreserveBackslash\centering}p{#1}}  \newcolumntype{R}[1]{>{\PreserveBackslash\raggedleft}p{#1}}  \newcolumntype{L}[1]{>{\PreserveBackslash\raggedright}p{#1}}

\ifCLASSINFOpdf
\else
\fi
\begin{document}

\title{Mobile Energy Transfer in Internet of Things}

\author{Qingqing~Zhang,~\IEEEmembership{Student Member,~IEEE},
Gang~Wang,~\IEEEmembership{Member,~IEEE},
Jie~Chen,~\IEEEmembership{Fellow,~IEEE},
Georgios~B.~Giannakis,~\IEEEmembership{Fellow,~IEEE},
and~Qingwen~Liu,~\IEEEmembership{Senior Member,~IEEE}

\thanks{The work of Q.~Zhang, and Q.~Liu was supported by the National Natural Science Foundation of China Grant 61771344. The work of G.~Wang and G.~B.~Giannakis was supported partially by the National Science Foundation under Grants 1514056, and 1711471. The work of J.~Chen was partially supported by the National Natural Science Foundation of China under Grants U1509215, and by the Program for Changjiang Scholars and Innovative Research Team in University (IRT1208). \emph{(Corresponding author: Qingwen Liu.)}}

\thanks{Q.~Zhang and Q.~Liu are with the College of Electronic and Information Engineering, Tongji University, Shanghai 201800, China.
	
	G.~Wang and G.~B.~Giannakis are with the Department of Electrical and Computer Engineering, University of Minnesota, Minneapolis, MN 55455, USA (e-mail: gangwang@umn.edu;  georgios@umn.edu).
	
	J.~Chen is with the State Key Lab of Intelligent Control and Decision of Complex Systems, School of Automation, Beijing Institute of Technology, Beijing 100081, China, and also with Tongji University, Shanghai 200092, China (e-mail:  chenjie@bit.edu.cn).}

}

\maketitle

\begin{abstract}
Internet of things (IoT) is powering up smart cities by connecting all kinds of electronic devices. The power supply problem of IoT devices constitutes a major challenge in current IoT development, due to the poor battery endurance as well as the troublesome cable deployment. The wireless power transfer (WPT) technology has recently emerged as a promising solution. Yet, existing WPT advances cannot support free and mobile charging like Wi-Fi communications. To this end, the concept of mobile energy transfer (MET) is proposed, which relies critically on an resonant beam charging (RBC) technology. The adaptive (A) RBC technology builds on RBC, but aims at improving the charging efficiency by charging devices at device preferred current and voltage levels adaptively. A mobile ARBC scheme is developed relying on an adaptive source power control.
%Moreover, \anne{a $1000$mAh Li-ion battery is taken as an example charging target.}
 Extensive numerical simulations using a $1,000$mAh Li-ion battery show that the mobile ARBC outperforms simple charging schemes such as the constant power charging, the profile-adaptive charging, and the distance-adaptive charging in saving energy.
\end{abstract}


\IEEEpeerreviewmaketitle

%%%%%%%%%%%%%%%%%%%%%%%%%%%%%%%%%%%%%%%%%%%%%%%%%%%%%%%%%%%%%%%%%%%%%%%%%%%%%%%%%%%%%%%%%%%%%%%%%%%%%%%%%%%%%%%%%%%%
\section{Introduction}\label{introduction}
Internet of things (IoT) is promising a smart and comfortable life by connecting IoT devices \cite{iotding,xiao2017iot,internetyu,iotcheng}. However, these devices such as smart-phones, laptops, actuators, and sensors are battery powered or wired \cite{aloi2017phone,chenjie2018sensor,isikdag2015computers,kato2017actuators,lazarescu2013sensor}. To charge IoT devices anytime anywhere, users may carry power cords or seek the power outlets, which brings them inconveniences \cite{lowpowerChen,phoneyu,mobilecheng,sensorling}. The power supply problem has become one of the critical challenges in IoT \cite{phoneyu,chenjie2017sensor}. Therefore, the wireless power transfer (WPT) technologies were advocated \cite{wpt,akan2018battery}.

Existing WPT technologies are mainly based on inductive coupling, magnetic resonance, and magnetic induction \cite{liu2016dlc}. Taking safety into consideration, these technologies only support charging devices with low power over short distances. Users cannot get their devices charged safely over long distances while in transit. To meet user-specific charging requirements, we propose the concept of mobile energy transfer (MET) in this paper. MET aims at providing a `Wi-Fi-like' charging service, enabling `safe anytime anywhere charging,' so users do not need to worry about the battery endurance.

\begin{figure}
	\centering
	\includegraphics[scale=0.38]{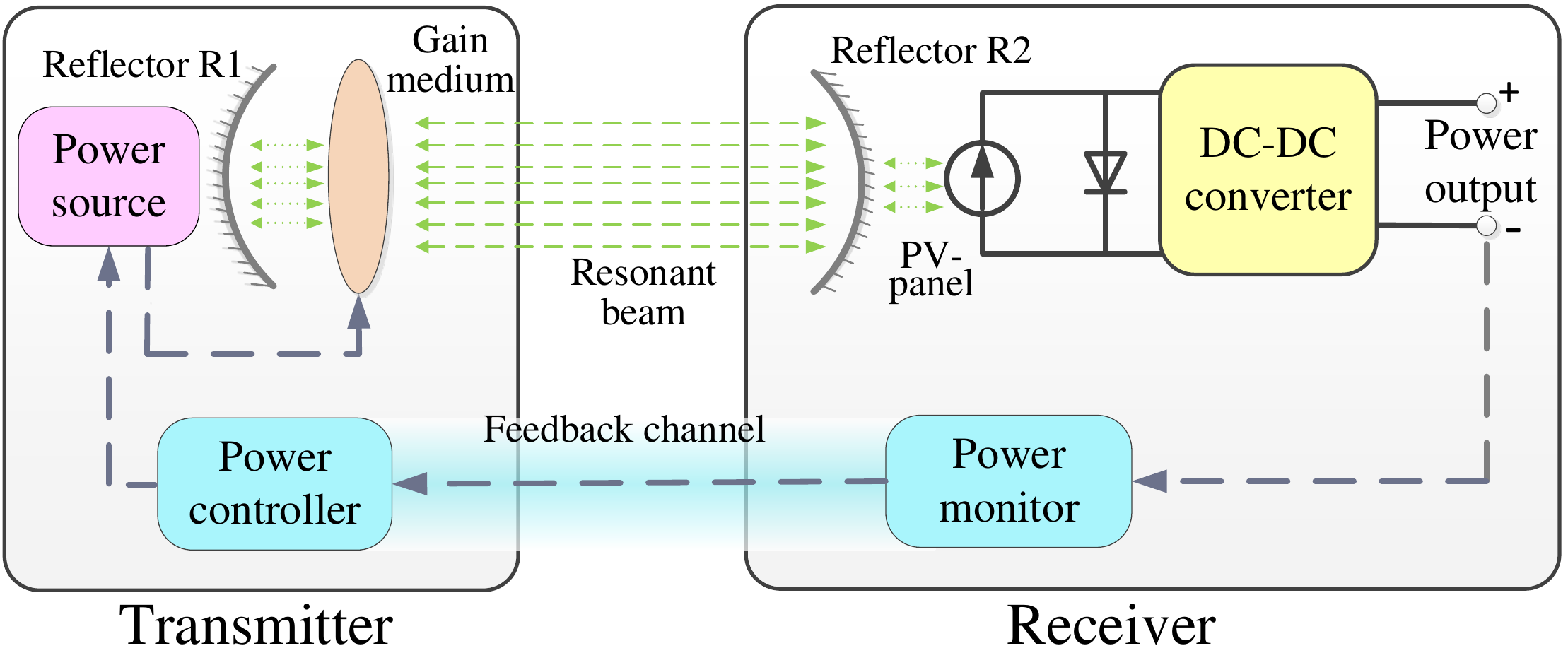}
	\caption{Standard ARBC system architecture. An ARBC system consists of two spatially separated parts: an ARBC transmitter and an ARBC receiver. The transmitter provides wireless power for the receiver, and the receiver can be integrated into the devices to be charged.}
	\label{adaptivestructure}
\end{figure}

As an MET technology, resonant beam charging (RBC) was introduced and developed in \cite{liu2016dlc,dlcqing}. To maximize the RBC efficiency, the adaptive (A) RBC was proposed in \cite{arbcqing}. See Fig.~\ref{adaptivestructure} for a standard ARBC system architecture, which consists of two spatially independent parts: an ARBC transmitter and an ARBC receiver. Compared with the RBC system, a power monitor and a power controller are augmented to support adaptive power control through feedback. On this basis, IoT devices equipped with ARBC receivers can be charged with battery preferred currents and voltages, hence charged with battery preferred power.

\begin{figure*}
	\centering
	\includegraphics[scale=0.24]{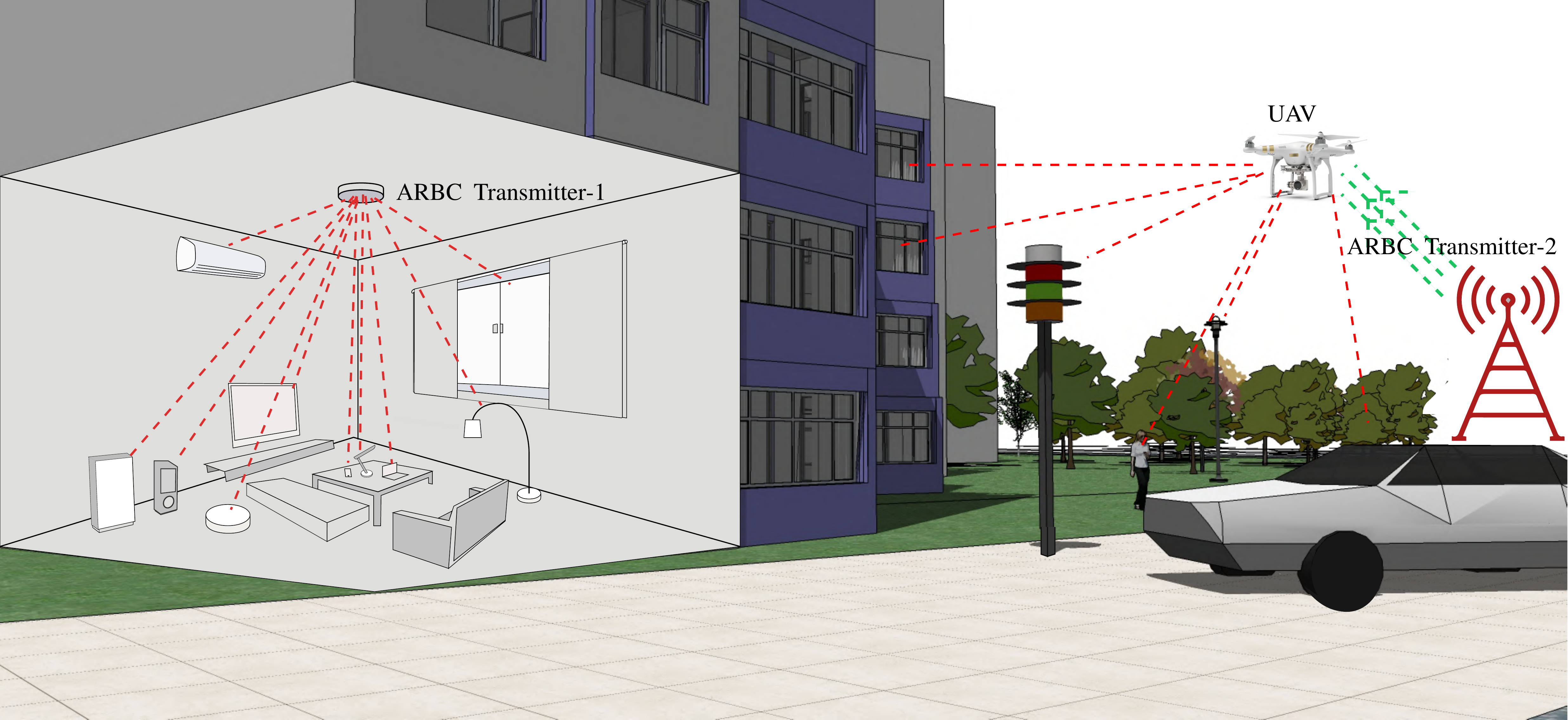}
	\caption{Representative ARBC application scenarios. Both indoor and outdoor charging examples are shown. All receivers within the coverage of the transmitter can be simultaneously charged. A UAV in an outdoor setting can be used as an ARBC relay if integrated with both a transmitter and a receiver.}
	\label{arbcnetwork}
\end{figure*}

The ARBC procedure developed in \cite{arbcqing} is outlined as follows: i) The power monitor at the receiver gets the output power, i.e., the battery preferred charging power, and feeds it back to the power controller at the transmitter; ii) The power controller informs the power source with the battery preferred charging value; iii) The power source provides the battery preferred power to the gain medium and stimulates out the resonant beam; iv) The resonant beam transmits through the air with some attenuation and arrives at the receiver; v) The beam power is converted to the laser power, and subsequently to the electric power; and, vi) After being converted to the battery preferred charging current and voltage by the DC-DC converter, the electric power can be used to charge the battery.

Figure~\ref{arbcnetwork} describes a representative ARBC application scenario. In this ARBC system, multiple receivers can be charged simultaneously by a single transmitter, so long as all receivers are placed within the coverage of the transmitter. In the indoor scenario, all electronic devices can be charged by Transmitter-1, which is installed in a ceiling light. In the outdoor setting, an unmanned aerial vehicles (UAV) is used as an ARBC relay \cite{uavding,uavcheng}. Within the coverage of Transmitter-2, the UAV can be charged by the Transmitter-2, whereas it can also work as a transmitter to charge devices within its coverage, such as the smart-phone and the street lamp.

The ARBC performance is evaluated by analyzing a Li-ion battery charging procedure; see \cite{arbcqing} for details. The Li-ion battery is charged from the empty state to the fully charged state over a certain charging distance, namely the charging distance between the transmitter and the receiver. However, in our daily life, the battery requesting for charging services can be at a random state of charge (SOC) rather than at the running-out state of charge. The charging distance is modeled time-varying during the charging procedure rather than treated as a constant \cite{dynamiczhang}. For different SOCs, the battery preferred charging values are different, which will bring different source power requirements. Moreover, for a certain source power, the received power falls off with the increment of the charging distance. Take the daily charging requirements into consideration, a mobile ARBC procedure, where the batteries at a random SOC while in transit, is simulated to illustrate MET.

In a nutshell, the contributions of the present paper can be summarized as follows:

c1) To charge IoT devices' batteries safely over a long distance in movement, we propose the concept of mobile energy transfer (MET);
	%, which relies critically on an RBC technology;} %\wang{should elaborate more on the key of the MET with one sentence; if I understand correctly, the battery-preferred... is the key making it different than WPT?}

c2) To realize MET, we develop a novel mobile ARBC scheme, which controls the source power at the transmitter adaptively such that the charging target is charged at battery preferred current and voltage levels; and,
% a closed-form adaptive source power control scheme is proposed. The source power control relies on the target output power of the ARBC system. To estimate the target power, a widely used Li-ion battery is taken as the charging target, and the battery preferred charging power is the target output power. The battery preferred charging power model} is put forward; and,

c3) To assess the MET performance, we perform substantial numerical {simulations} on a randomly moving IoT device's Li-ion battery.
% \anne{Both constant power charging (CPC), profile-adaptive charging (PAC), distance-adaptive charging (DAC), and ARBC can contribute to realizing MET.
	Performance analysis suggests that ARBC could save at least $35.1\%$ source energy relative to other simple charging schemes.
%	about $69.3\%$, $54.6\%$, and $35.1\%$ source energy relative to CPC, PAC, and DAC, respectively.

In the rest of this paper, the MET mechanism of ARBC is detailed using a dynamically moving Li-ion battery in Section~\ref{Smechanism}. The MET performance is evaluated through numerical simulations on the mobile Li-ion battery in Section~\ref{Sperformance}. Finally, the paper is concluded in Section~\ref{conclusions}, along with the outlook for future research.

%%%%%%%%%%%%%%%%%%%%%%%%%%%%%%%%%%%%%%%%%%%%%%%%%%%%%%%%%%%%%%%%%%%%%%%%%%%%%%%%%%%%%%%%%%%%%%%%%%%%%%%%%%%%%%%%%%%%%
%%%%%%%%%%%%%%%%%%%%%%%%%%%%%%%%%%%%%%%%%%%%%%%%%%%%%%%%%%%%%%%%%%%%%%%%%%%%%%%%%%%%%%%%%%%%%%%%%%%%%%%%%%%%%%%%%%%%%

\section{Mobile Energy Transfer}\label{Smechanism}
RBC is a promising technology to realize WPT, which is of great importance for wireless sensor network\cite{wsnChen}. ARBC, which is based on RBC, can improve the charging efficiency. To enable mobile ARBC, the source power is controlled to supply the battery with its preferred charging current and voltage in real time. A novel mobile ARBC scheme is developed with an adaptive source power control. The source power relies on the target output power of the ARBC system and the relative charging distance between the transmitter and the receiver. Commonly used in most IoT devices, the Li-ion battery is taken as the charging target in this paper. The target output power is the battery preferred charging power, which is obtained numerically by fitting the Li-ion battery charging profile on real data. Moreover, the charging distance is modeled time-varying during the entire charging procedure. Finally, we summarize the MET process of an ARBC system.

\subsection{Mobile ARBC Scheme}\label{}
In an RBC system, IoT devices are charged with the target output power $P_{o}$, which depends on the source electric power $P_{s}$ and the resonant beam transmission efficiency $\eta_{t}$, yet $\eta_{t}$ relies on the structure of the RBC system and the charging distance between the transmitter and the receiver. The end-to-end power relationship, namely the relationship between $P_{o}$ and $P_{s}$, can be given by
\begin{equation}\label{poutpinww}
P_{o} = a \eta_{t} P_{s}+b
\end{equation}
where $a$ and $b$ are constant coefficients determined by the structure of the RBC system (e.g., the size and reflectivity of the reflectors, and parameters of the PV-panel, to name a few).

To enhance the energy efficiency of RBC, an initial concept of ARBC was proposed in our precursors \cite{arbcqing,dlcqing}, by introducing adaptive source power control into RBC through a feedback channel. In these previous works, only the output power control is analyzed by simulating the battery charging over different distances. However, the key to realizing ARBC lies in adaptively controlling the source power $P_{s}$ to meet the battery preferred output power $P_{o}$.

From \eqref{poutpinww}, $P_{s}$ is determined by $P_{o}$, the charging distance $d$, and the other constant coefficients.  We can conclude that (e.g., \cite{dlcqing,rbcww,arbcqing})
\begin{equation}\label{pin-d-pout}
P_{s} = \frac{(P_{o}-m)(1+f)}{2n(1-f)}\left({e^{ - 2\pi \frac{{{r^2}}}{{\lambda (l+d)}}}} - \ln f\right)
\end{equation}
where $m$, $f$, $n$, and $r$ are some constant coefficients. Specifically, $m$ depends on the ARBC system structure, $n$ is related with the material and the structure of the system and the PV-panel, $f$ denotes the reflectivity of reflector R2, %the resonate cavity and the PV-panel
and $r$ is the radius of R1 and R2; $\lambda$ captures the resonant beam wavelength, $l$ refers to the distance between the gain medium and the reflector R1. For a certain ARBC system structure, $\lambda$ and $l$ are constants. $d$ is the distance between the gain medium and the reflector R2, i.e., {the charging distance between the transmitter and the receiver}. If the receiver keeps changing its relative position to the transmitter in the ARBC system, $d$ may take different values.

\begin{figure}
	\centering
	\includegraphics[scale=0.58]{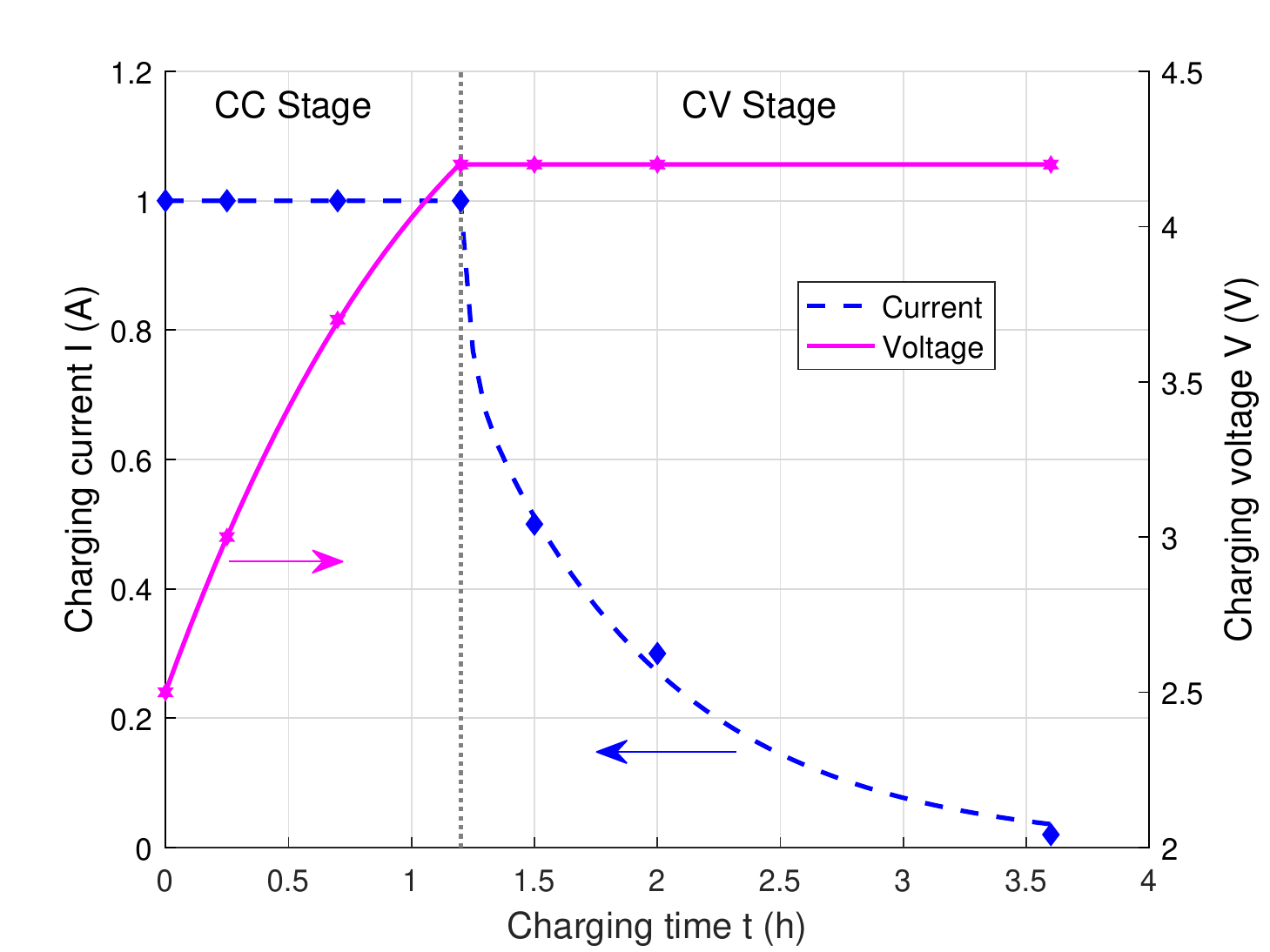}
	\caption{The Li-ion battery charging current and voltage profile. Both the current and voltage vary with the charging duration. At the beginning, the charging current is constant while the voltage increases with the charging time. After some period, the current drops sharply but the voltage is saturated.}
	\label{chargingprofile}
\end{figure}

\subsection{{Target Output Power}}\label{}
{With regards to the source power $P_{s}$ control, the value of the target output power $P_{o}$ should be available. For an IoT device, the target output power is the device's battery preferred charging power.} For different types of batteries, {the preferred charging power $P_{o}$ is different.} Even for batteries of the same type, $P_{o}$ varies with the SOCs of batteries. Among all IoT devices, smart-phones are necessities in our daily life. Moreover, most smart-phones are Li-ion battery powered, which motivates well taking a $1,000$ mAh Li-ion battery as the charging target. The Li-ion battery charging profile, which shows how the preferred charging current and voltage change with the charging time, can be divided into $2$ stages: the constant current (CC) stage and the constant voltage (CV) stage \cite{battery2stage,batterydata}.
%For different {SOCs}, the corresponding charging time is different, and the preferred charging values are hence different \cite{wenfair}.

\begin{figure}
	\centering
	\includegraphics[scale=0.58]{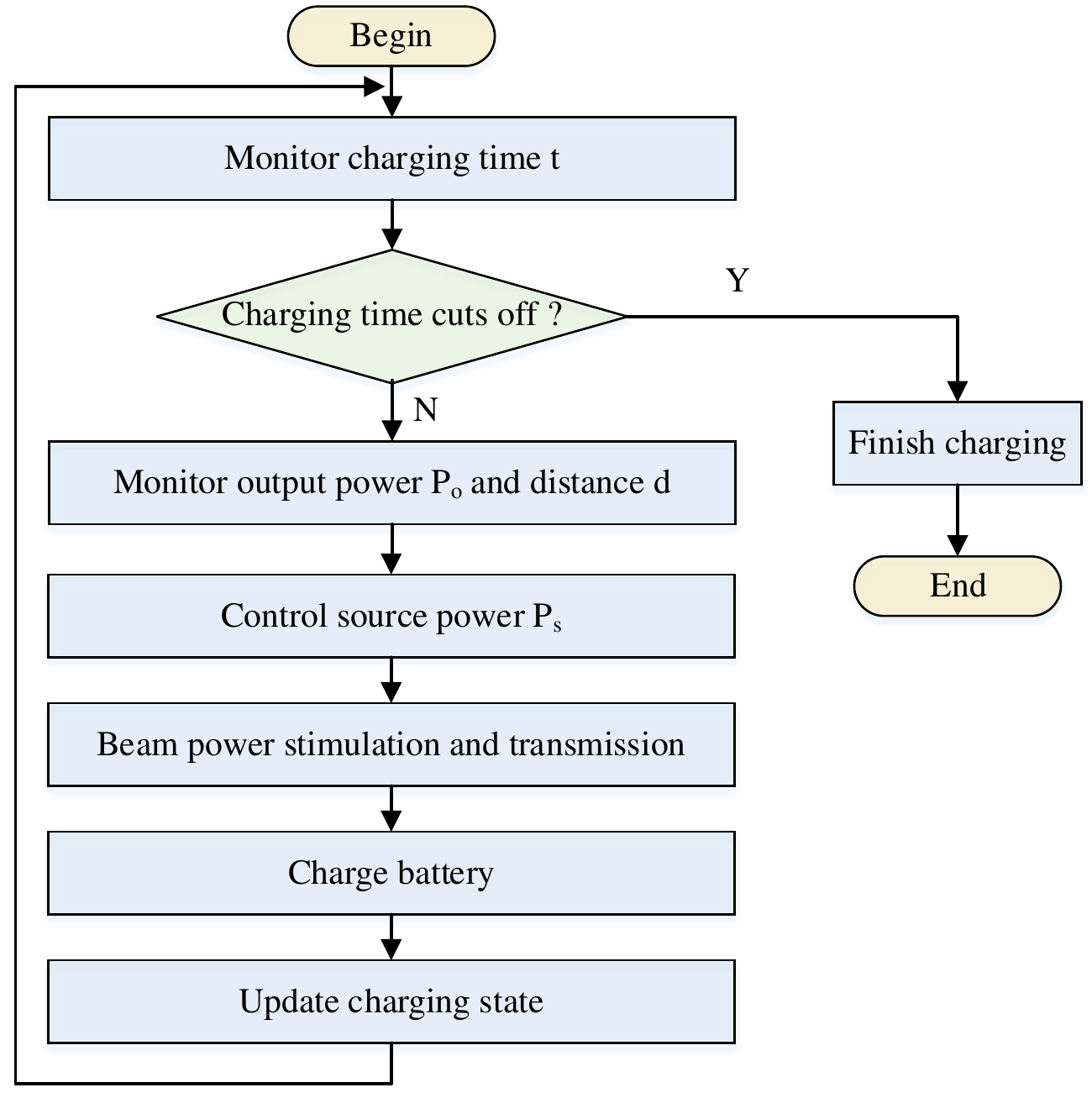}
	\caption{The MET Flow in ARBC. The charging time of battery is monitored at the receiver. If the time is not cut off, the battery will keep being charged, and the SOC will keep changing. Otherwise, if the charging time cuts off, the charging procedure is finished.}
	\label{adaptiveflow}
\end{figure}

To charge the Li-ion battery adaptively, the preferred charging power should be computed in real time. To this end, the preferred charging current and voltage should be obtained at first. A 2-stage Li-ion battery charging profile was described in \cite{battery2stage,batterydata,arbcqing}. We propose the following model to capture the charging current $I$ and voltage $V$ over time
\begin{equation}\label{chargingcv}
\begin{aligned}
I= \left\{
             \begin{array}{ll}
             1.0,&0\le t<1.2  \\
             a_{i}e^{b_{i} t} + c_{i}e^{d_{i} t},& 1.2\le t<3.6
             \end{array}
\right.,
% and \quad
\\[1mm]
V= \left\{
             \begin{array}{ll}
             a_{v}e^{b_{v}t} + c_{v}e^{d_{v}t}, & 0\le t<1.2 \\
             4.2, & 1.2\le t<3.6
             \end{array}
\right.
\end{aligned}
\end{equation}
where $t$ denotes the charging time, $a_{i}$, $b_{i}$, $c_{i}$, $d_{i}$, and $a_{v}$, $b_{v}$, $c_{v}$, $d_{v}$ are some coefficients determined by characteristics of the battery.

In Fig.~\ref{chargingprofile}, stars and diamonds depict the simulated battery charging data from \cite{arbcqing}. The dash line shows the variation of charging current $I$, while the solid one that of charging voltage $V$ obtained from \eqref{chargingcv}. As can be seen from the curves, our model \eqref{chargingcv} agrees with {the charging profile in \cite{arbcqing}}.

Multiplying $I$ and $V$ in \eqref{chargingcv}, the battery preferred charging power, namely the target output power $P_{o}$, can be defined as
\begin{equation}\label{chargingp}
P_{o}=IV=  \left\{
             \begin{array}{ll}
             a_{v}e^{b_{v}t} + c_{v}e^{d_{v}t}, & 0\le t< 1.2  \\
             4.2(a_{i}e^{b_{i} t} + c_{i}e^{d_{i} t}), & 1.2\le t<3.6.
             \end{array}
\right.
\end{equation}
Using \eqref{chargingp}, $P_{o}$ can be obtained for any given charging time slot. For a certain SOC, the charging time instant can be determined; hence, the battery preferred charging values can be obtained.

\begin{figure}%[t]
	\centering
	\includegraphics[scale=0.58]{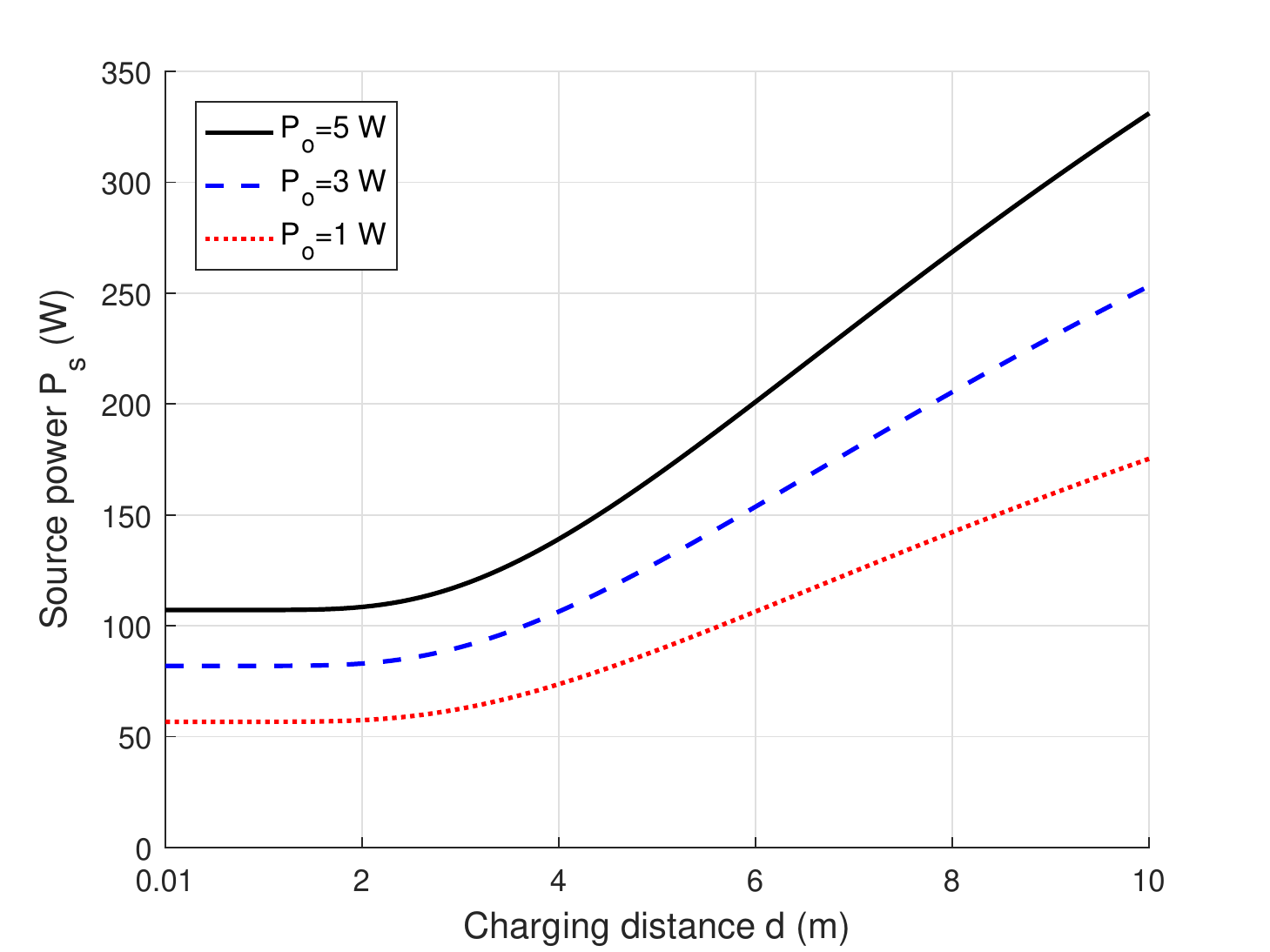}
	\caption{Required source power $P_{s}$ as a function of charging distance $d$ to produce constant output power $P_{o}$ (1, 3, and 5W).}
	\label{pind}
\end{figure}

\subsection{{Mobile Energy Transfer Process}}\label{}
When the ARBC system structure is fixed, the coefficients in \eqref{pin-d-pout},  \eqref{chargingcv}, and \eqref{chargingp} can be determined.
% \wang{In reality, a given battery requiring for charging may be at any SOC.
%Therefore, the preferred output power $P_{o}$ may take different values. this sentence has been repeated at least twice...} Moreover, the position of the battery relative to the transmitter, that is the charging distance $d$ may be dynamically changing during the charging process.
According to \eqref{pin-d-pout}, if both $d$ and $P_{o}$ are given, $P_{s}$ can be computed to provide the basis for the adaptive control in ARBC. {The MET} process of ARBC is depicted in Fig.~\ref{adaptiveflow}, with its main steps summarized as follows.
%\begin{enumerate}%[(a)]

  S1) The power monitor tracks the SOC of the battery, and the charging time $t$;

  S2) If the charging time is cut off, the charging process terminates; otherwise, turn to step S3);

  S3) The power monitor obtains the preferred charging power $P_{o}$, and sends it back to the power controller through the feedback channel;

  S4) The power controller obtains the distance of the receiver $d$ to the transmitter;

  S5) The power controller computes the source power $P_{s}$ based on $P_{o}$ and $d$, and requires $P_{s}$ from the power source;

  S6) The power source stimulates the gain medium, and generates resonant beam, which transfers through the air to the receiver;

  S7) The beam power is converted into electric power at the receiver. After being converted into the battery preferred current and voltage by a DC-DC convertor, the electric power is used to charge the battery; and,

  S8) The SOC of the battery and the charging time $t$ are updated. Turn to step S1).
%\end{enumerate}

With this MET process, the battery can be charged {at the battery preferred charging current} and voltage values, hence power even in transit during the charging procedure.
%The adaptive MET procedure can be completed.

%%%%%%%%%%%%%%%%%%%%%%%%%%%%%%%%%%%%%%%%%%%%%%%%%%%%%%%%%%%%%%%%%%%%%%%%%%%%%%%%%%%%%%%%%%%%%%%%%%%%%%%%%%%%%%%%%%%%
%%%%%%%%%%%%%%%%%%%%%%%%%%%%%%%%%%%%%%%%%%%%%%%%%%%%%%%%%%%%%%%%%%%%%%%%%%%%%%%%%%%%%%%%%%%%%%%%%%%%%%%%%%%%%%%%%%%%

\section{Performance Evaluation}\label{Sperformance}
The empirical performance of MET is assessed in this section. By default, only one receiver can be charged by a single transmitter at the same time. The constant coefficients in \eqref{pin-d-pout} can be readily obtained from \cite{rbcww,vmjpv,solarcell}, which are listed in Table~\ref{pindparameters}. All experiments were performed using MATLAB.

\begin{table}[b]%[symbol]
\renewcommand\arraystretch{1.3}
	\centering
	\caption{Power Conversion Parameter}
	\begin{tabular}{C{1.3cm} C{2cm}}
		\hline
		\textbf{Parameter} & \textbf{Value}  \\
		\hline
		\bfseries$m$ & -3.5017 \\
		\bfseries$n$ & 0.0795 \\
		\bfseries$f$ & 0.88 \\
		\bfseries$r$ & 1.5mm \\
		\bfseries$\lambda$ & 1064nm \\
		\bfseries$l$ & 65mm \\
		\hline
		\label{pindparameters}
	\end{tabular}
\end{table}

\begin{figure}%[b]
	\centering
	\includegraphics[scale=0.58]{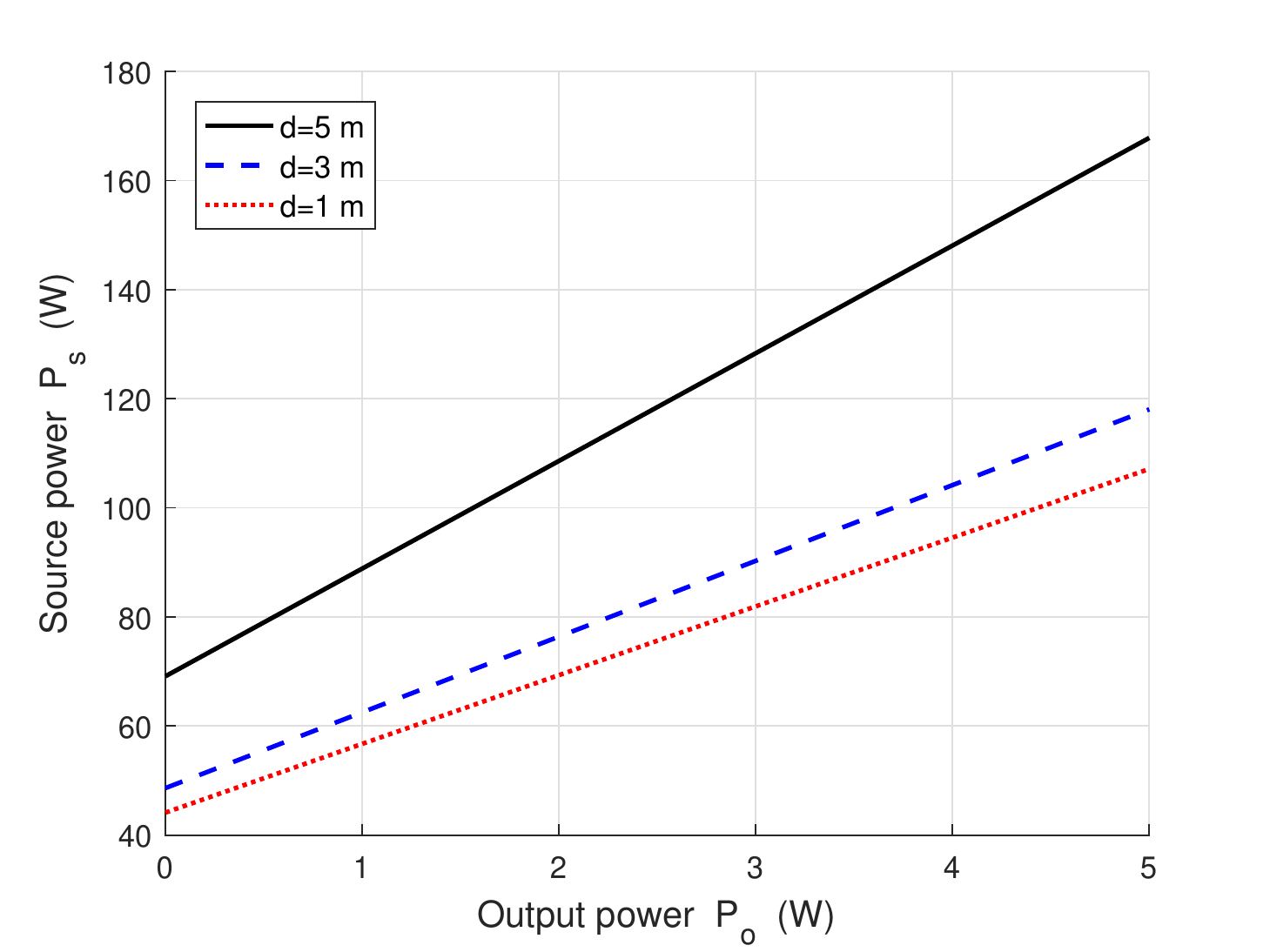}
	\caption{Required source power $P_{s}$ as a function of output power $P_{o}$ over constant charging distance $d$ (1, 3, and 5m).}
	\label{pinpout}
\end{figure}

\subsection{ARBC Performance}\label{}

\begin{table}[b]%[symbol]
\renewcommand\arraystretch{1.3}
\centering
\caption{Charging Profile Parameter}
\begin{tabular}{C{1.3cm} C{2cm}}
\hline
 \textbf{Parameter} & \textbf{Value}  \\
\hline
\bfseries$a_{i}$ & 3.4 \\
\bfseries$b_{i}$ & -1.263 \\
\bfseries$c_{i}$ & 1.873$\times10^{13}$ \\
\bfseries$d_{i}$ & -26.61 \\
\bfseries$a_{v}$ & 168.4 \\
\bfseries$b_{v}$ & -0.2903 \\
\bfseries$c_{v}$ & -165.9 \\
\bfseries$d_{v}$ & -0.3078 \\
\hline
\label{batparameters}
\end{tabular}
\end{table}

From \eqref{pin-d-pout}, it is clear that {$P_s$ depends on both $P_{o}$ and $d$.} When $P_{o}$ is given, we can simulate how $P_{s}$ changes with $d$, which is depicted in Fig.~\ref{pind} for $P_{o}=1$W, $3$W, and $5$W.
Plots show that when transmitting power over a short distance, say e.g., $d\le 2$m, the required source power $P_{s}$ remains almost a constant for fixed $P_o$. As $d$ increases, $P_{s}$ grows quickly. To provide a desired $P_{o}$ to the receiver, the required $P_{s}$ grows with $d$. For example, to charge the battery with $P_o=3$W over a distance of $2$m, an amount of $80$W or so source power is needed; if over a distance of $6$m, about $150$W source power is required. To charge a battery over $6$m, the required source power is at least about $200$W to provide with $5$W battery preferred power; it only takes slightly more than $100$W source power to provide $1$W battery preferred power over the same distance. Figure~\ref{pind} provides us with guidelines for source power and charging distance control to yield a desired output power at the receiver side.

On the other hand, for a fixed distance $d$, source power $P_{s}$ scales linearly with $P_{o}$; see Fig.~\ref{pinpout}. The dotted, dashed, and solid lines  show this linear relationship for $d=1$m, $3$m, and $5$m, respectively. According to Fig.~\ref{pind}, we find that $P_{s}$ admits the slowest increase when $d=1$m, while it grows at the fastest pace when $d=5$m. This agrees with the fact that the slope of the solid line is larger than that of the dotted line in Fig.~\ref{pinpout}. To obtain the preferred output power over a certain distance, Fig.~\ref{pinpout} offers guidelines for source power adjustment.

\begin{figure}
	\centering
	\includegraphics[scale=0.58]{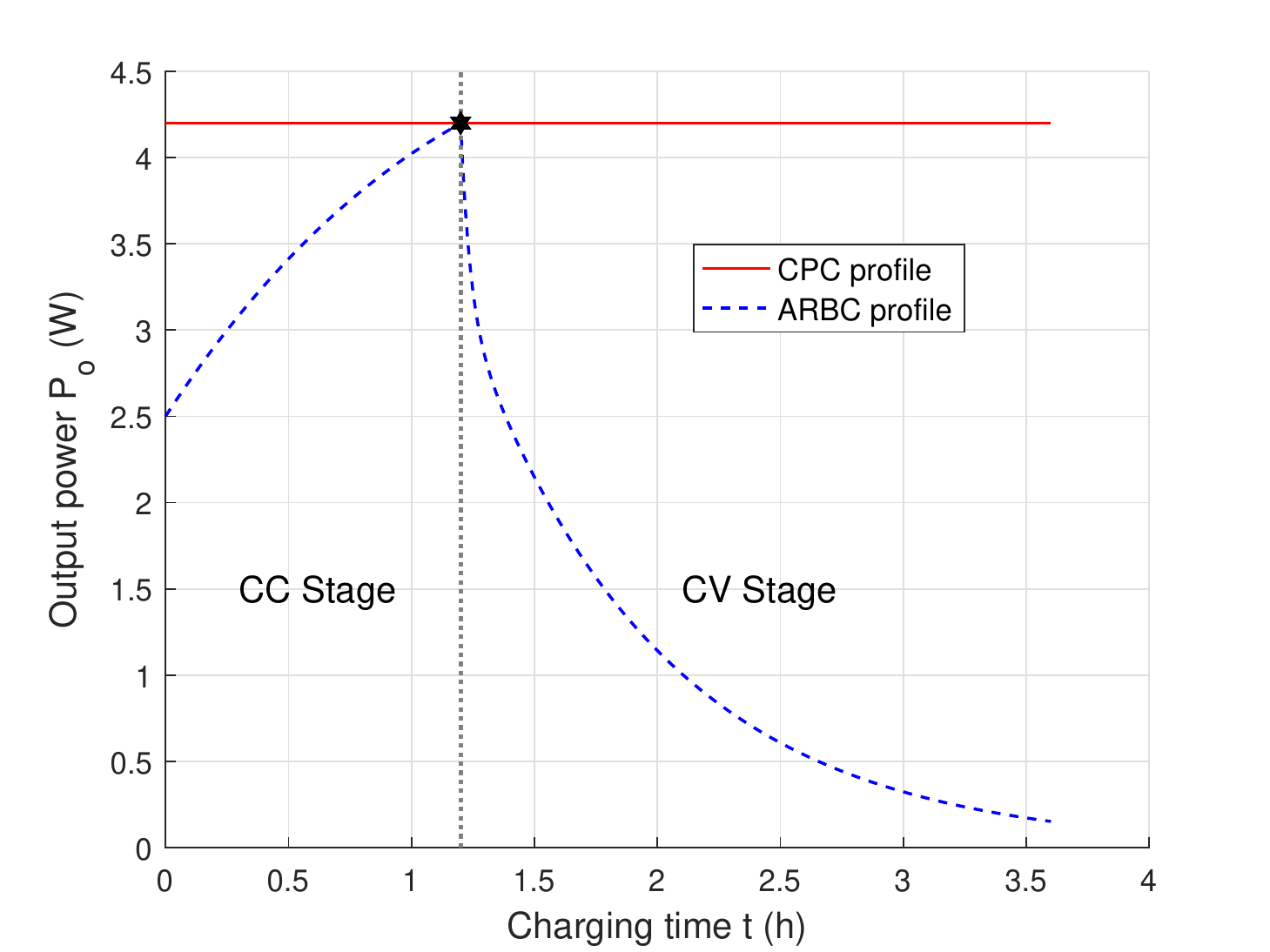}
	\caption{The Li-ion battery charging power profile of CPC and ARBC. That is, the changing characteristics of voltage-current set values during the battery charging period.}
	\label{powerprofile}
\end{figure}

\begin{figure}%[b]
	\centering
	\includegraphics[scale=0.4]{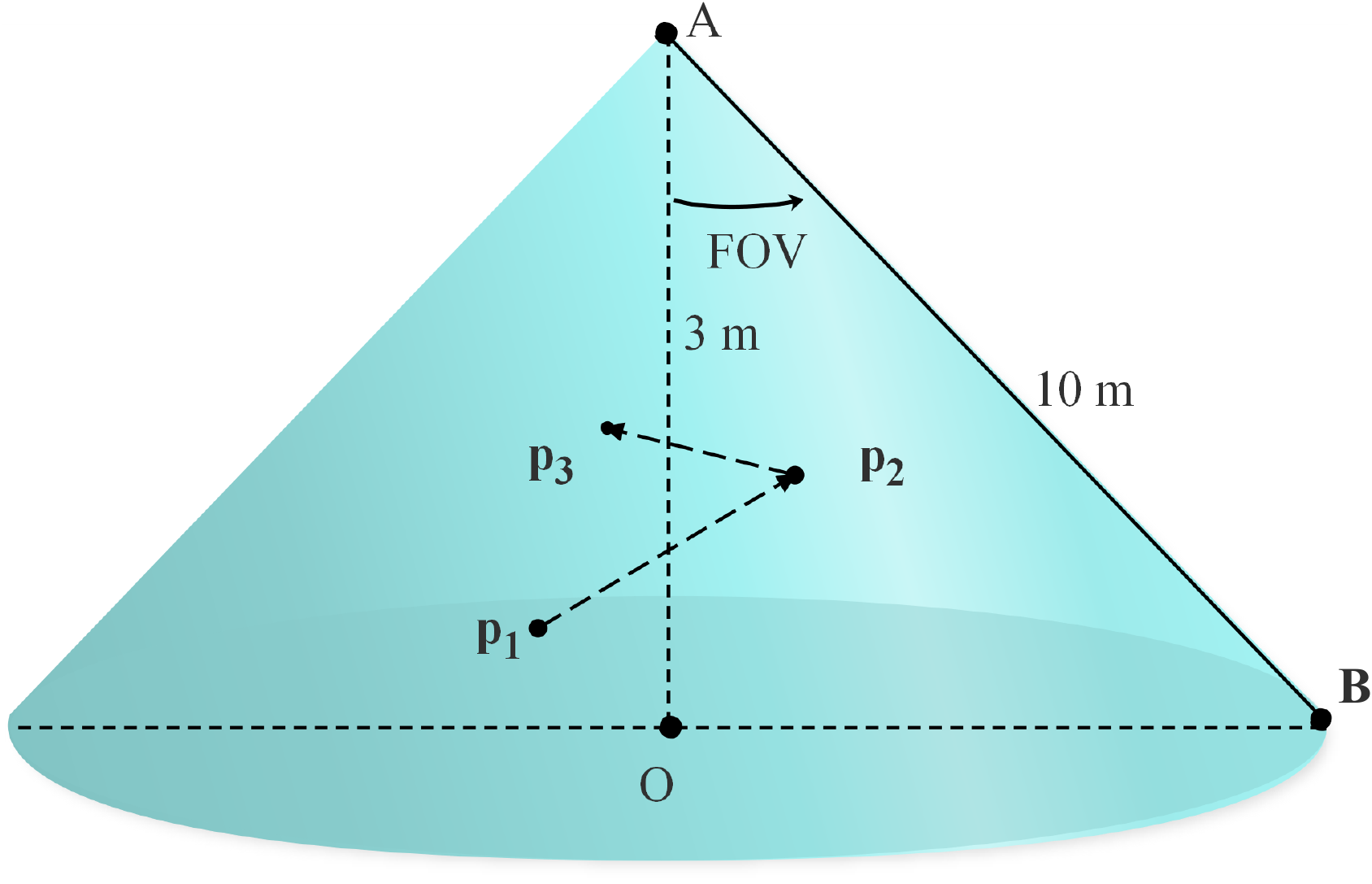}
	\caption{The ARBC transmitter's coverage. The transmitter is placed at point $A$; the farthest charging point (e.g., point $B$) is at the circumference of the flat circle, which is centered at point $O$.}
	\label{coverage}
\end{figure}

\begin{table}[b]%[symbol]
\renewcommand\arraystretch{1.3}
\centering
\caption{Charging Schemes}
\begin{tabular}{C{2.5cm} C{4.0cm}}
\hline
 \textbf{Scheme Acronym} & \textbf{Scheme Description}  \\
\hline
\bfseries CPC & Constant-power charging \\
\bfseries PAC & Profile-adaptive charging  \\
\bfseries DAC & Distance-adaptive charging \\
\bfseries ARBC & Adaptive resonant beam charging \\
\hline
\label{methodslist}
\end{tabular}
\end{table}

\subsection{Charging Power Profile}\label{}

\begin{figure}
	\centering
	\includegraphics[scale=0.58]{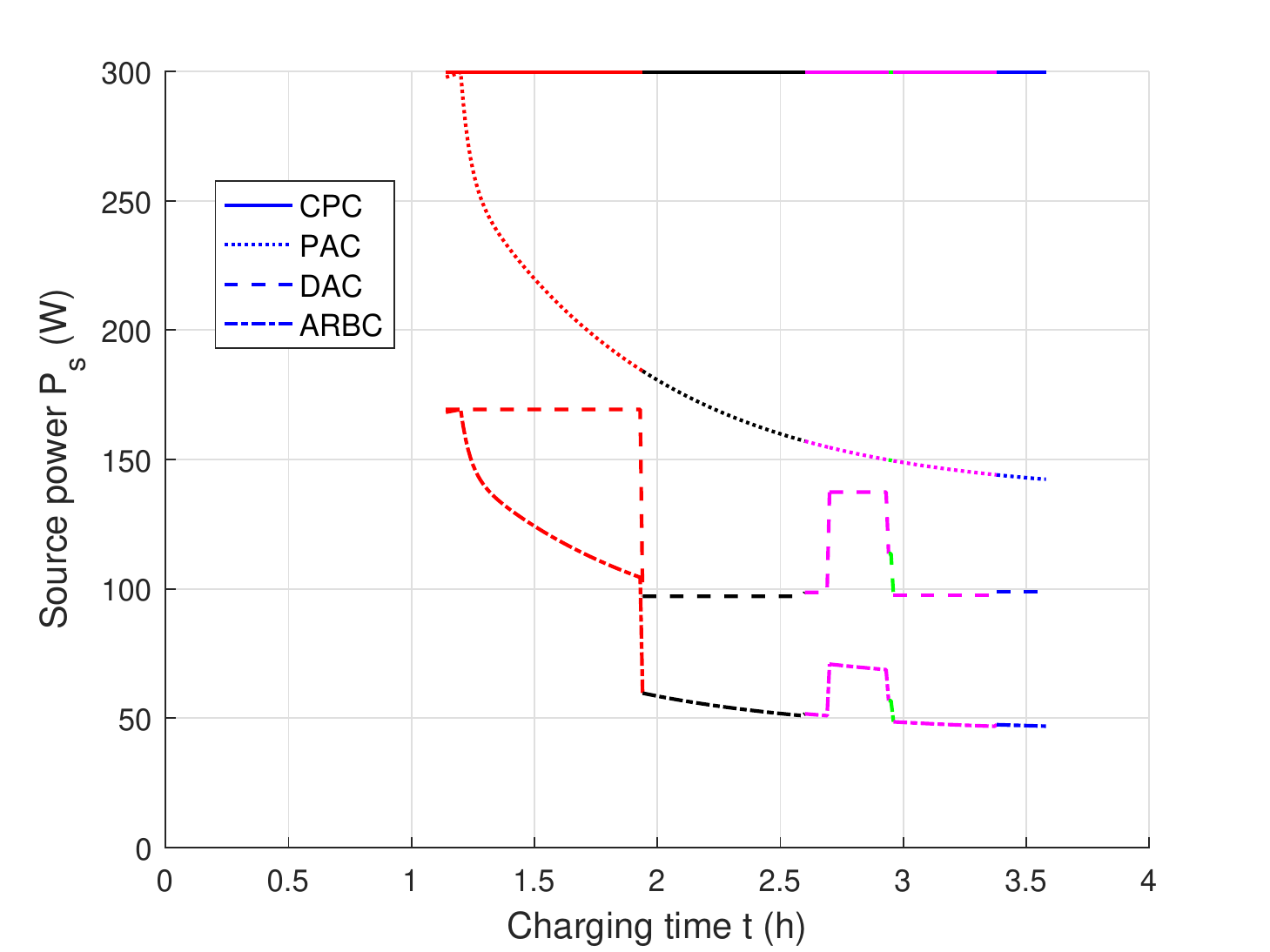}
	\caption{The source power consumed by different MET schemes. The preferred output power and the charging distance may be dynamically and randomly changing with the charging duration in a charging process.}
	\label{procedurepower}
\end{figure}

For a single cell Li-ion battery with $1,000$mAh capacity, the parameters in model \eqref{chargingcv}, or also in \eqref{chargingp}, take values listed in Table~\ref{batparameters}. Compliant with \eqref{chargingp}, the profile of battery preferred charging power, i.e., the battery charging power profile, is divided into the CC stage and the CV stage; see the dashed line in Fig.~\ref{powerprofile}. The battery preferred charging power is also the optimal output power $P_{o}$ of the {ARBC system}. At the CC stage, $P_{o}$ increases as the charging time passes. This is because the charging voltage increases while the charging current is kept fixed. Moreover, $P_{o}$ attains its maximum at the end of the CC stage; see the star marker in Fig.~\ref{powerprofile}. The CC stage is followed by the CV stage, in which $P_{o}$ decreases quickly due to the decrease of current yet with a constant voltage.  %\gang{For a certain SOC, the charging time instant can be determined,} \wang{have repeated this a couple of times, but not sure what it means and whether its necessary?} and the corresponding $P_{o}$ could be obtained according to the battery preferred charging power profile shown in Fig.~\ref{powerprofile}.

As a baseline, we consider a constant-power charging (CPC) scheme, which refers to that the battery preferred charging power keeps a constant power during the whole charging procedure. To meet the power demand at each time instant during the procedure, the constant value should be greater than or equal to the peak power in the charging procedure. The peak power, which  is $4.2$W ($1.0$A$\times$$4.2$V), is attained at the end of the CC stage marked by the star in Fig.~\ref{powerprofile}. {The solid line depicts the profile of CPC.}

To summarize, Fig.~\ref{powerprofile} describes the charging profiles of a Li-ion battery using CPC and ARBC from the under-charged state, which refers to the start of the CC stage, to the fully charged state, i.e., the end of the CV stage.

\subsection{{Mobile Energy Transfer Performance}}\label{}
In a daily charging scenario, the battery of the device requested for charging services can be at any SOC. Moreover, the charging distance, namely the relative position between the transmitter and the receiver, may be also changing during the entire charging procedure. That is, as long as the receiver is placed within the transmitter's coverage, the charging distance can be time-varying and take certain random values.

\begin{figure}
	\centering
%\bigskip
%\medskip
%\addvspace {3pt}
\setlength{\parskip}{1em}
	\includegraphics[scale=0.58]{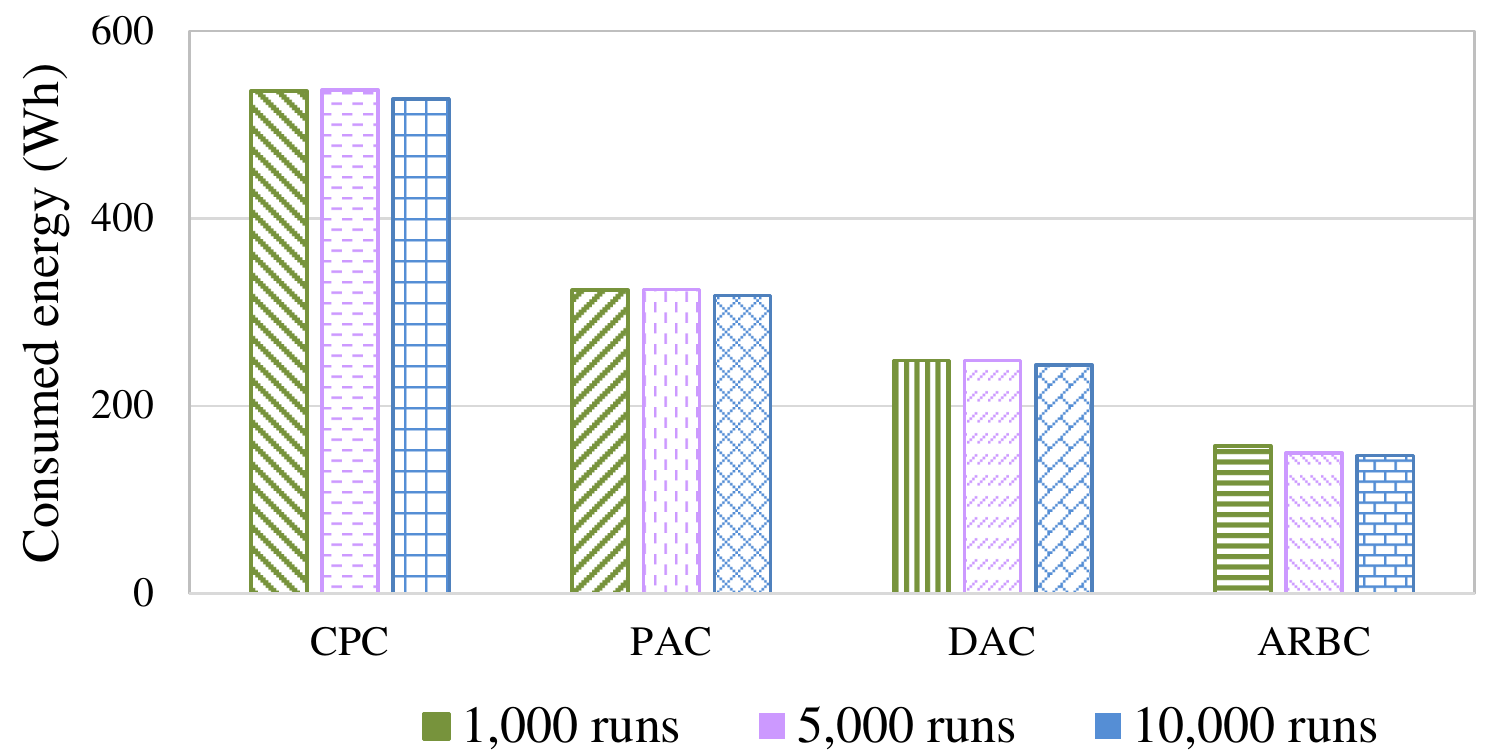}
	\caption{Average energy consumption of CPC, PAC, DAC, and ARBC schemes after running 1,000, 5,000, and 10,000 times.}
	\label{averconsum}
\end{figure}

To evaluate the MET performance in a more realistic environment, we simulate the mobile charging process by giving a mobile Li-ion battery charging instance within a certain coverage. To benchmark MET, we assume that the maximum charging distance is $10$m, and the maximum height is $3$m; see Fig.~\ref{coverage} for the coverage of the ARBC transmitter in our considered setup. The transmitter is placed at point $A$; the farthest charging point (e.g., point $B$) is at the circumference of the flat circle, which is centered at point $O$ with a radius of $\sqrt{91}$. Given the charging distance and the maximum height, the field of view (FOV) of the ARBC system is determined. As can be seen, the spatial coverage of the transmitter is in a cone shape. The receiver's battery can be charged from any position within the coverage of the transmitter.

In our MET example, a charging procedure consists of multiple charging periods. In the first period, the battery is charged at a random SOC, and it can be placed at a random location e.g., point $p_{1}$ in Fig.~\ref{coverage}. After being charged for a certain period, the battery randomly changes its position say moving to point $p_{2}$, and the second charging period begins. If the charging procedure is not completed, the battery may move to another place such as $p_{3}$, so on and so forth. The charging procedure will be terminated when it is interrupted or the battery is fully charged.

Since the device often moves quickly from one position to another, the time for realizing this movement is ignored here for simplicity. Hence, the end time of a charging period is just the beginning of the next one. Therefore, the initial SOC in one period, but not the first period, is just the ending SOC in the previous one. For a certain charging period, the relative distance takes a certain value. But for all the periods, the distance may take different value randomly. Moreover, we assume that the battery is charged without any interruption until it is fully charged.
%accidental termination or the

\begin{figure}
	\centering
	\includegraphics[scale=0.7]{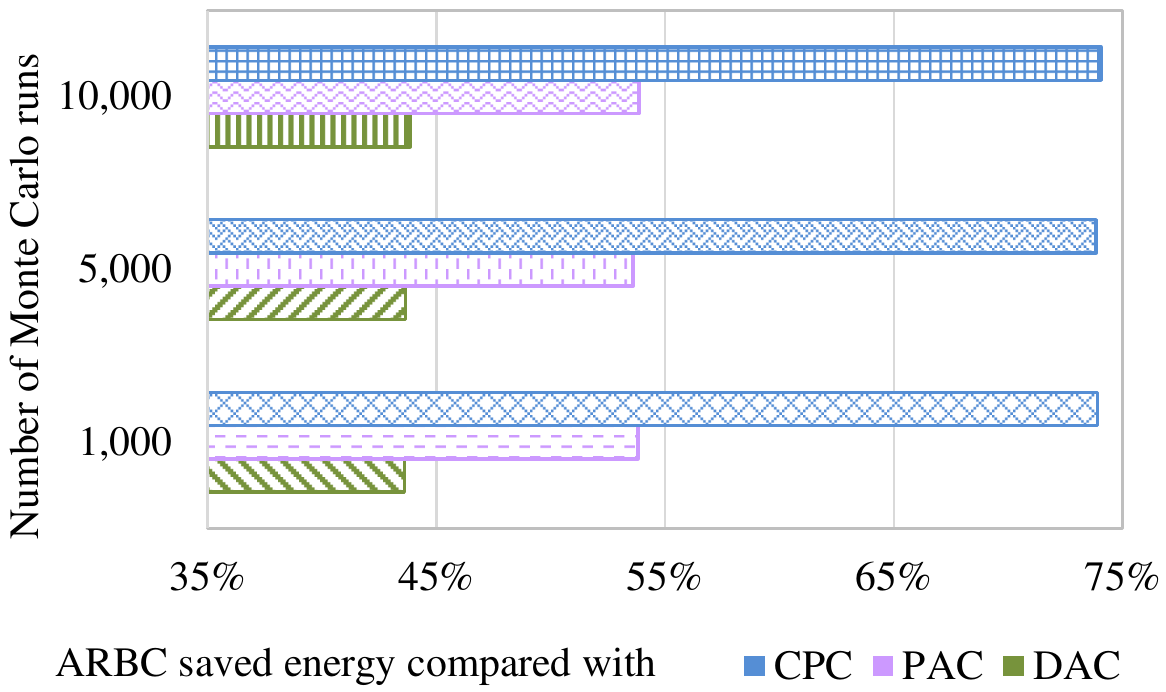}
	\caption{Average percentage of energy saved by ARBC compared with DAC, PAC, and CPC.}
	\label{aversave}
\end{figure}

To assess the MET performance, in addition to CPC, the profile-adaptive charging (PAC) and distance-adaptive charging (DAC),  were also simulated as baselines. The PAC scheme supports charging the battery taking into consideration the battery charging power profile over a certain distance, while the DAC scheme prefers charging the battery with a fixed power over randomly changing distances. The four charging schemes (CPC, PAC, DAC, and ARBC) are listed in Table~\ref{methodslist}.

For CPC, the battery preferred charging power is equal to the peak preferred charging power, as the solid line shows in Fig.~\ref{powerprofile}. To be able to provide charging services for any device within the coverage, the charging distance is set to be $10$m, the maximum charging distance that the transmitter can cover. Hence, using \eqref{pin-d-pout}, the CPC required source power can be computed. See the solid line in Fig.~\ref{procedurepower} for the source power variation of CPC during the charging procedure. It is clear that the solid line remains constant over time.

For PAC, the battery preferred charging power is time-varying with the distance taking the maximum value. The dot line in Fig.~\ref{procedurepower} shows the PAC preferred source power variation during the charging procedure, which follows the power-profile trend.

For DAC, the battery is charged with the peak preferred power while in transit during the charging procedure. See Fig.~\ref{procedurepower}, the dash line depicts the DAC preferred source power changing trend during the charging procedure. As can be seen, the source power keeps the same during a certain charging period. {At different time slots, the charging distance may be changed, and the preferred source power may vary correspondingly.}

For ARBC, both the charging profile and the charging distance are taken into consideration when determining the required amount of source power for output. The dash-dot line plots the source power variation of ARBC during the charging procedure in Fig.~\ref{procedurepower}. It is clearly shown that the charging distance takes the same value at a certain time slot, so the dash-dot line maintains the charging power profile trend. The charging distance at one time slot differs from the others.

From Fig.~\ref{procedurepower}, in a certain charging time instant, CPC requires the largest amount of source power, followed by PAC and DAC, and ARBC requires the least power. However, only one exemplary charging procedure is taken into consideration in our MET instance. %our MET instance is just an exemplary application.
In daily charging scenarios, the initial SOC of the being charged battery, the time slot, and the charging distance may take random values. To evaluate the MET performance %analyze how much energy will be consumed by the four charging methods on average
in daily scenarios, we simulate the aforementioned mobile Li-ion battery charging instance for $1,000$, $5,000$, and $10,000$ times.

Figure~\ref{averconsum} shows the average energy consumed by the four schemes during the charging process in our simulation. Clearly, the ARBC scheme consumes the least amount of energy on average compared with the other charging schemes. For a fixed charging scheme, the amount of energy consumed after averaging over $1,000$, $5,000$, and $10,000$ Monte Carlo runs remains almost the same. The average energy consumption of CRC is about $530$Wh, and that for PAC, DAC, and ARBC is about $320$Wh,  $248$Wh, and $150$Wh, respectively.

See Fig.~\ref{aversave} for the average energy saving (in percentage) of ARBC relative to other charging schemes. The average energy saving achieved by the proposed charging scheme maintains almost the same after simulating the charging procedure for $1,000$, $5,000$, and $10,000$ times. ARBC can save about $74.0\%$, $53.8\%$, and $43.7\%$ source energy when compared with CRC, PAC, and DAC, respectively.

%%%%%%%%%%%%%%%%%%%%%%%%%%%%%%%%%%%%%%%%%%%%%%%%%%%%%%%%%%%%%%%%%%%%%%%%%%%%%%%%%%%%%%%%%%%%%%%%%%%%%%%%%%%%%%%%%%%%

\section{Conclusions}\label{conclusions}

The concept of ``mobile energy transfer (MET)'' was proposed in this paper. To implement MET, the source power is controlled in an adaptive manner, aiming to provide battery preferred charging power. The MET mechanism was presented using the mobile Li-ion battery as an example, while the SOC of the battery as well as the relative distance between the transmitter and battery can take random values. Numerical analysis show that about $74.0\%$, $53.8\%$, and $43.7\%$ energy is saved by ARBC relative to CPC, PAC, and DAC, respectively.

The present work also opens up several interesting directions for future research, which include evaluating the charging schemes using commonly used batteries on prototype ARBC systems, and investigating the limit of transmitting power.

%\section{Acknowledgment}\label{acknowledge}

%The authors would like to thank Hao Deng, Mingliang Xiong and Yunfeng Bai in Tongji University for their valuable suggestions. We also thank Xiaoyan Liang and Wei Wang in the Shanghai Institute of Optics and Fine Mechanics, Chinese Academy of Sciences, for their help in improving the energy transmission model of ARBC.

\bibliographystyle{IEEEtran}
\bibliographystyle{unsrt}
\bibliography{references}

\vspace{-13 mm}
\begin{IEEEbiography}[{\includegraphics[width=1in,height=1.25in,clip,keepaspectratio]{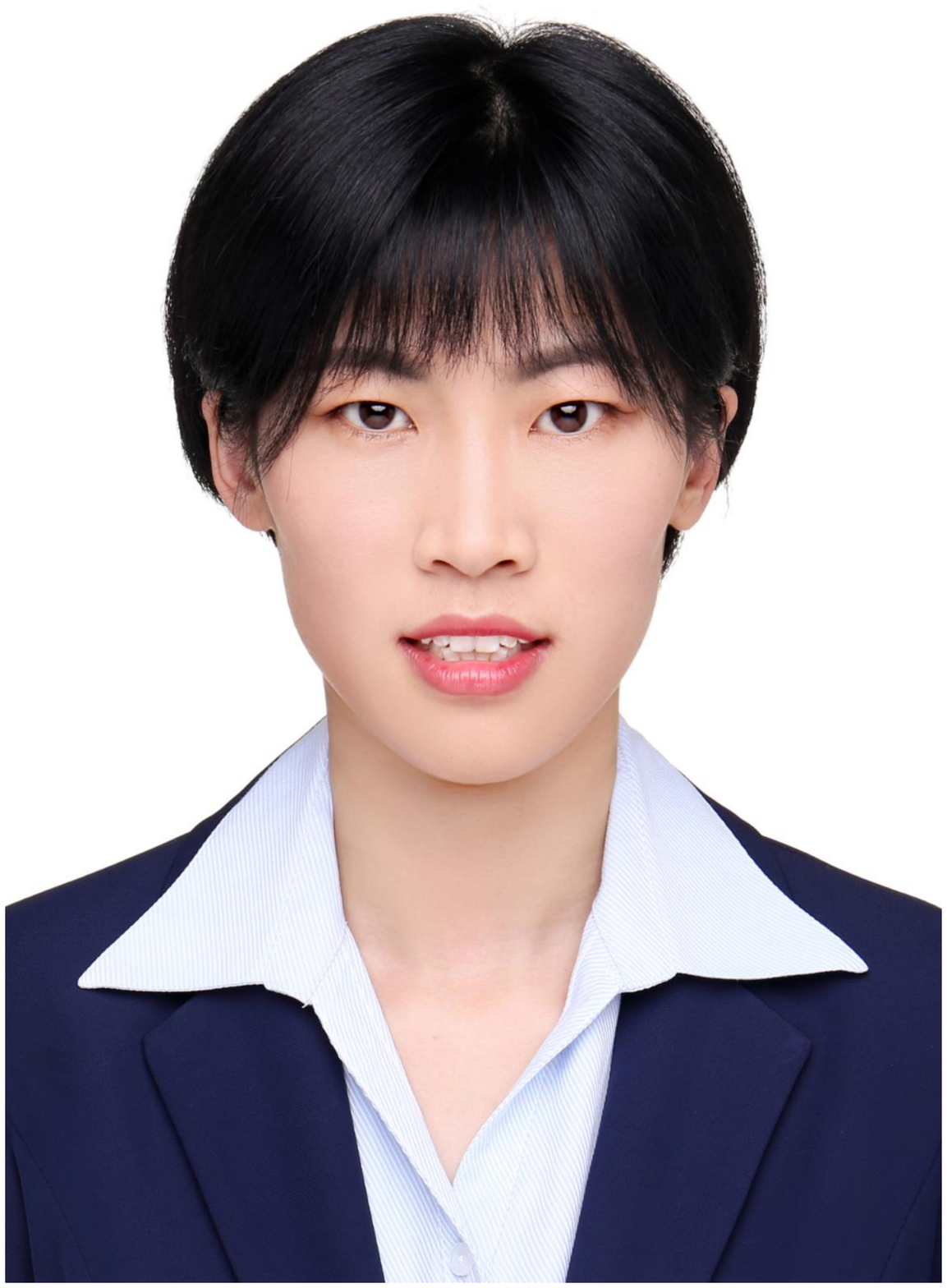}}] {Qingqing Zhang} (S'16) received the M.Sc. degree in computer system and architecture from the Hunan University of Technology, Hunan, China, in 2016. She is currently working toward the Ph.D. degree in the College of Electronics and Information Engineering, Tongji University, Shanghai, China.
Her research interests include wireless power transfer, wireless communications, and Internet of Things.
\end{IEEEbiography}
%\vspace{-5 mm}
\begin{IEEEbiography}[{\includegraphics[width=1in,height=1.25in,clip,keepaspectratio]{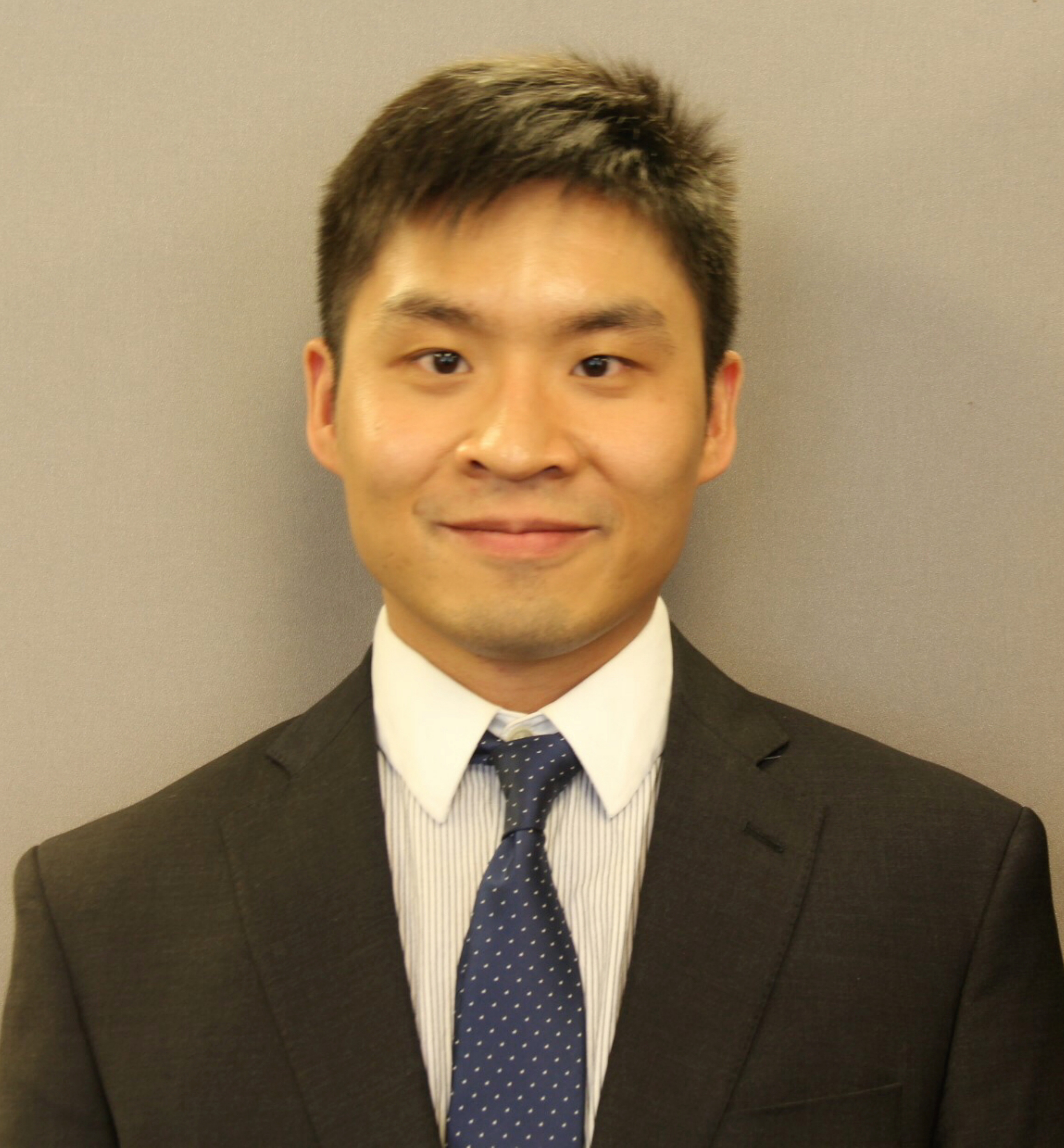}}] {Gang Wang} (M'18) received his B.Eng. in electrical engineering and automation from the Beijing Institute of Technology, Beijing, China, in 2011, and his Ph.D. in electrical and computer engineering from the University of Minnesota, Minneapolis, MN, USA, in 2018. He is currently a postdoctoral associate in the Department of Electrical and Computer Engineering at the University of Minnesota.
	
His research interests focus on the areas of statistical signal processing, optimization, and deep learning with applications to data science and smart grids. He received Paper Awards at the 2017 European Signal Processing Conference and the 2019 IEEE Power \& Energy Society General Meeting.
\end{IEEEbiography}
%\vspace{-13 mm}
\begin{IEEEbiography}[{\includegraphics[width=1in,height=1.25in,clip,keepaspectratio]{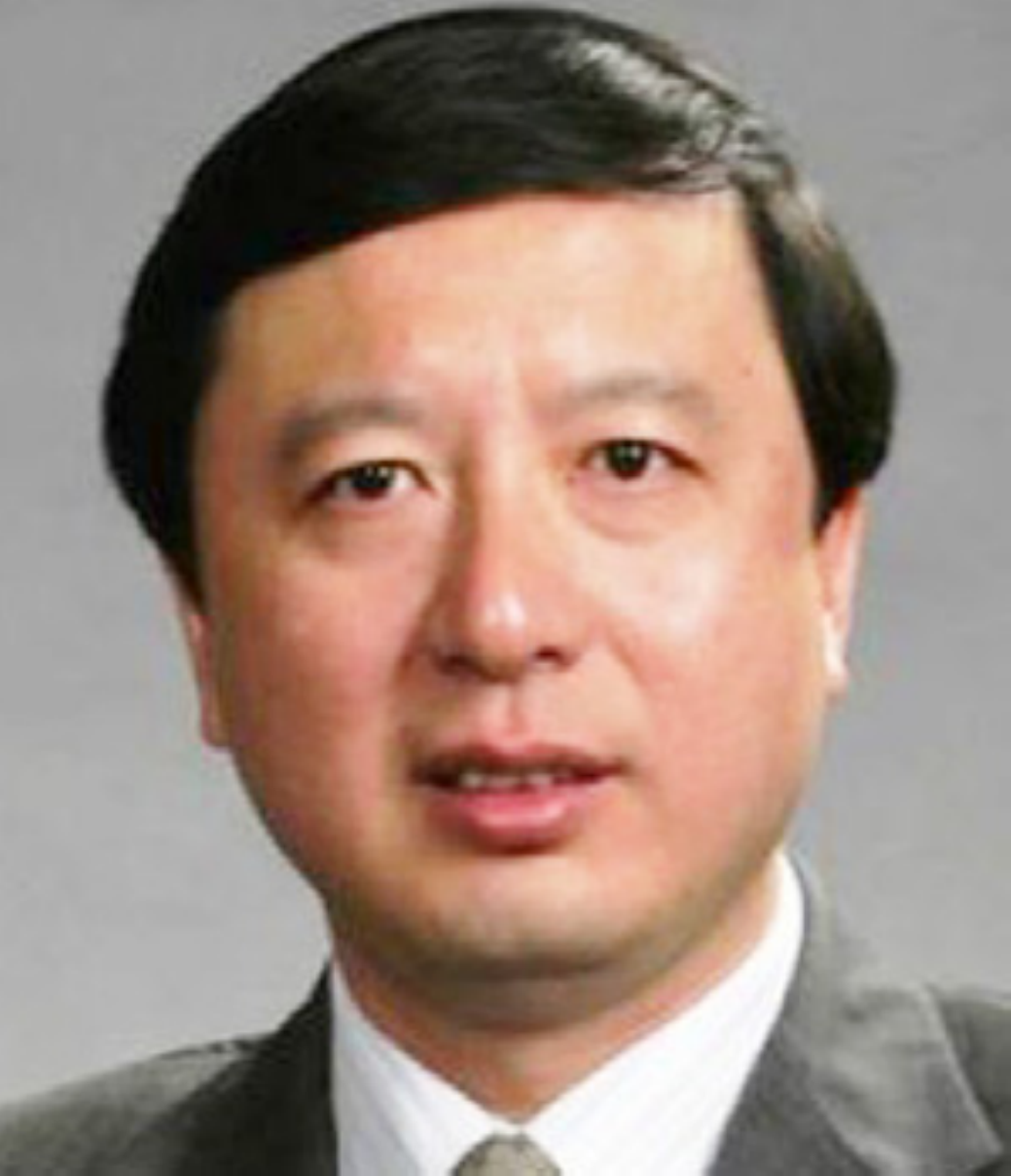}}]{Jie Chen} (F'19) received his B.Sc., M.Sc., and the Ph.D. degrees in control theory and control engineering from the Beijing Institute of Technology, Beijing, China, in 1986, 1996, and 2001, respectively. From 1989 to 1990, he was a visiting scholar at the California State University, Long Beach, California, USA. From 1996 to 1997, he was a research fellow in the School of Engineering at the University of Birmingham, Birmingham, UK. Since 1986, he has been with the Beijing Institute of Technology, where he is now a professor in control science and engineering, and serves the head of the State Key Laboratory of Intelligent Control and Decision of Complex Systems, China.
	
His main research interests include intelligent control and decision in complex systems, multi-agent systems, nonlinear control, and optimization. He has (co-)authored 4 books and more than 200 research papers. He served as a managing editor for the \emph{Journal of Systems Science \& Complexity}, and an associate/subject editor for several other international journals, including the \emph{IEEE Transactions on Cybernetics}, and the \emph{International Journal of Robust and Nonlinear Control}. He is a member of the Chinese Academy of Engineering.
\end{IEEEbiography}

%\vspace{-13 mm}
\begin{IEEEbiography} [{\includegraphics[width=1in,height=1.25in,clip,keepaspectratio]{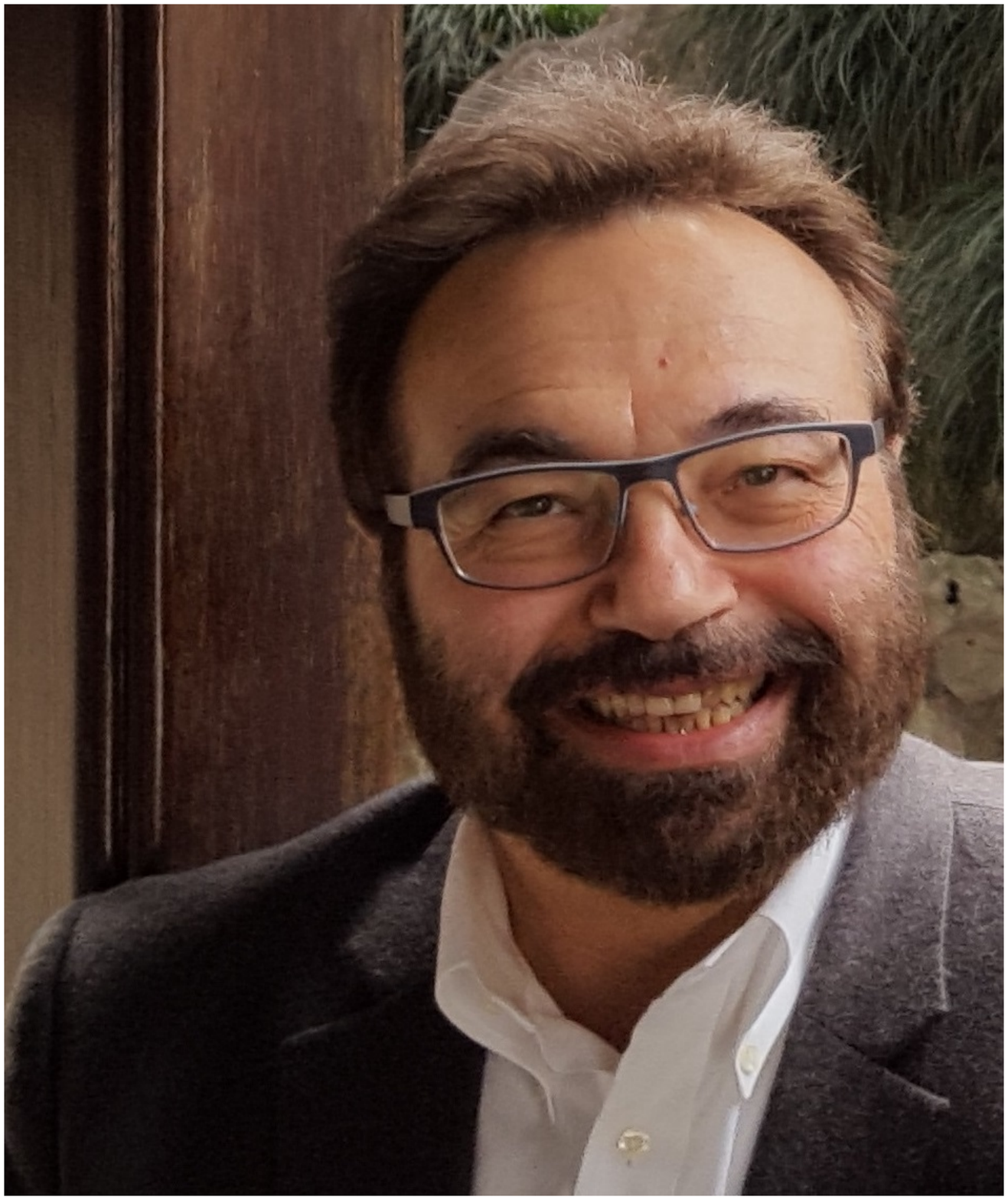}}] {Georgios B. Giannakis} (F'97) received his Diploma in Electrical Engr.. from the Ntl. Tech. Univ. of Athens, Greece, 1981. From 1982 to 1986 he was with the Univ. of Southern California (USC), where he received his MSc. in Electrical Engr., 1983, MSc. in Mathematics, 1986, and Ph.D. in Electrical Engr., 1986. He was a faculty member with the University of Virginia from 1987 to 1998, and since 1999 he has been a professor with the Univ. of Minnesota, where he holds an ADC Endowed Chair, a University of Minnesota McKnight Presidential Chair in ECE, and serves as director of the Digital Technology Center.
	
His general interests span the areas of statistical learning, communications, and networking - subjects on which he has published more than 450 journal papers, 750 conference papers, 25 book chapters, two edited books and two research monographs (h-index 142). Current research focuses on Data Science, and Network Science with applications to the Internet of Things, social, brain, and power networks with renewables. He is the (co-) inventor of 32 patents issued, 	and the (co-) recipient of 9 best journal paper awards from the IEEE Signal Processing (SP) and Communications Societies, including the G. Marconi Prize Paper Award in Wireless Communications. He also received Technical Achievement Awards from the SP Society (2000), from EURASIP (2005), a Young Faculty Teaching Award, the G. W. Taylor Award for Distinguished Research from the University of Minnesota, and the IEEE Fourier Technical Field Award (inaugural recipient in 2015). He is a Fellow of EURASIP, and has served the IEEE in a number of posts, including that of a Distinguished Lecturer for the IEEE-SPS.
\end{IEEEbiography}
%\vspace{-13 mm}

\begin{IEEEbiography}[{\includegraphics[width=1in,height=1.25in,clip,keepaspectratio]{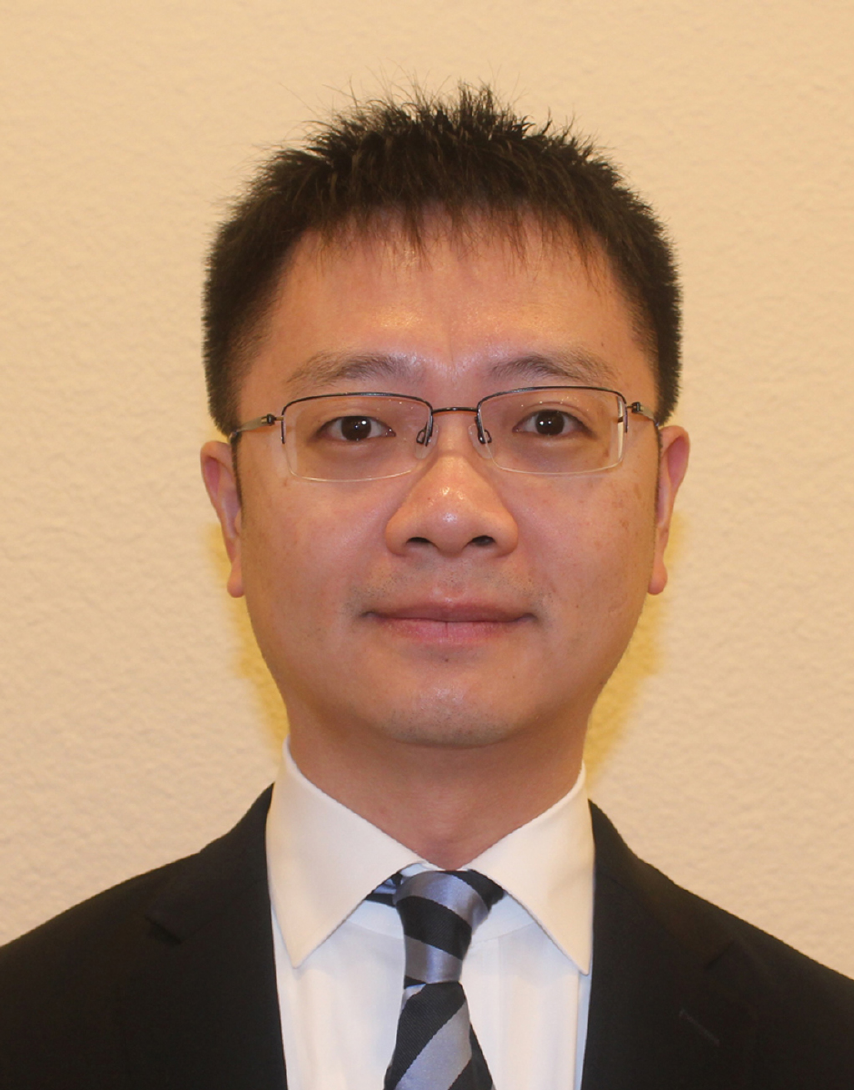}}]{Qingwen Liu} (M'07-SM'15) received his B.S. degree in electrical engineering and information science from the University of Science and Technology of China, Hefei, in 2001 and his M.S. and Ph.D. degrees from the Department of Electrical and Computer Engineering, University of Minnesota, Minneapolis, in 2003 and 2006, respectively. He is currently a professor with the College of Electronics and Information Engineering, Tongji University, Shanghai, China.
His research interests lie in the areas of wireless power transfer and Internet of Things.
\end{IEEEbiography}

\end{document}